\documentclass[a4paper,11pt]{article}
\pdfoutput=1 

\usepackage{jcappub} 

\usepackage[T1]{fontenc} 

\title{\boldmath Equatorial Periodic Orbits and Gravitational Wave Phenomenology around Spherically-symmetric vacuum solution in Freund-Nambu scalar-tensor gravity }

\author[a,b]{Dhruba Jyoti Gogoi}
\author[c,d]{Jyatsnasree Bora}
\author[e]{Himanshu Chaudhary}
\author[f]{M. Yousaf}
\author[g]{G. Mustafa}

\affiliation[a]{Department of Physics, Madhabdev University, Narayanpur, Lakhimpur 784164, Assam, India}
\affiliation[b]{Research Center of Astrophysics and Cosmology, Khazar University, Baku, AZ1096, 41 Mehseti Street, Azerbaijan}

\affiliation[c]{Department of Physics, Madhabdev University, Narayanpur, Lakhimpur 784164, Assam, India}
\affiliation[d]{Pacif Institute of Cosmology and Selfology (PICS) Sagara, Sambalpur 768224, Odisha, India}  

\affiliation[e]{Department of Physics, Babeș-Bolyai University, Kogălniceanu Street, Cluj-Napoca, 400084, Romania}

\affiliation[f]{Department of Mathematics, Virtual University of Pakistan, 54-Lawrence Road, Lahore 54000, Pakistan}

\affiliation[g]{Department of Physics, Zhejiang Normal University,
Jinhua 321004, China}
   
\emailAdd{moloydhruba@yahoo.in}
\emailAdd{jyatnasree.borah@gmail.com}
\emailAdd{himanshu.chaudhary@ubbcluj.ro}
\emailAdd{myousaf.math@gmail.com}
\emailAdd{gmustafa3828@gmail.com}

\abstract{We investigate test particle dynamics and gravitational wave (GW) phenomenology in an exact spherically symmetric vacuum solution of Freund-Nambu scalar-tensor gravity. This framework generalizes the Janis-Newman-Winicour (JNW) naked singularity via a geometric non-linear coupling $q$ and a direct scalar-particle coupling $g_s$. We demonstrate that these parameters systematically modify the Innermost Stable Circular Orbit (ISCO)—which shifts inward for $g_s > 0$—and the Marginally Bound Orbit (MBO). Furthermore, we classify bound periodic trajectories to isolate extreme zoom-whirl orbits exhibiting intense periapsis precession. By applying the Numerical Kludge method to Extreme Mass-Ratio Inspirals (EMRIs), we reveal that scalar-tensor corrections induce a macroscopic temporal dephasing in high-frequency GW bursts, even when the orbit's spatial topology is preserved. These unique phase shifts offer a robust diagnostic signature for future space-based observatories like LISA to probe the strong-field regime and constrain scalar-tensor extensions of general relativity.}


\begin{document}\maketitle\flushbottom

\section{Introduction}

Einstein's formulation of general theory of relativity (GR) presents gravity as a geometric description of spacetime, while at its core, the framework is constructed upon the metric tensor, which characterizes the gravitational field. This is one of the basic reason due to which the theory is often referred to as a tensor based formulation of gravity, whereas prior to Einstein's work, attempts made to describe gravitation through scalar approaches, notably by G. Nordstrom, who extended the Newtonian potential into a relativistic scalar quantity. However, such scalar formulations lacked a true geometric interpretation, consequently, the equivalence principle, which constitutes one of the fundamental pillars of GR, was not naturally incorporated within these early models during the initial developments. Einstein was not satisfied with this limitation and ultimately developed a dynamical scenario in which the geometry of spacetime itself evolves in response to matter and energy, although this formulation initially appeared speculative. Later it confirmed through extensive observational evidence spanning various physical phenomena. Moreover, the theory served as a theoretical model which illustrated how innovative conceptual scenarios can transition into well established physical descriptions, despite its success and its recognition as the standard theory of gravitation. GR also motivated the development of alternative theories. Scalar tensor theories attracted significant attention among these alternative theories. From a preliminary perspective, such theories might seem like a revival of earlier scalar gravity models. However, these alternative theories differ fundamentally, rather than merely combining scalar and tensor fields. Scalar tensor theories constructed upon the good foundation of GR. In these scenarios scalar field contributes in a nontrivial manner, typically through non minimal coupling with the geometric sector whereas the origin of such type of theories can be traced back to the work of Jordan in which the embedding of a four dimensional curved manifold within a higher dimensional flat spacetime explored where additional degree of freedom emerged as a scalar field. These additional degree of freedom can be interpreted as allowing the gravitational constant to vary across spacetime and idea aligns with Dirac's hypothesis and such a perspective extends beyond the conventional scope of the standard gravitational scenario and higher dimensional theories proposed in literature \cite{bailin1987kaluza,gladush2003five,FujiiMaeda2003}. These approaches aimed to unify gravity with other fundamental interactions. Building upon these foundational ideas, scalar tensor theories evolved into a rich class of models and these are regarded as compelling extensions of the standard $\Lambda$CDM cosmological scenario~\cite{BransDicke1961,FujiiMaeda2003,SotiriouFaraoni2010}. However, such interactions may lead to distinct physical effects and measurable deviations from GR predictions~\cite{Doneva:2022ewd,Berti2015,Barack2019}.

An early realization of such type of gravity is provided by the Freund-Nambu theory which originally proposed by Freund and his collaborators in~\cite{Freund1968PR} as well as this formulation extends GR. This extension introduced a dynamical scalar field that contributes to gravitational interactions in conjunction with the spacetime metric. The presence of this additional field enriches the dynamical behavior of the theory and can significantly influence both cosmological evolution and the internal structure of compact entities. Within the Freund-Nambu scenario, the gravitational action is generalized to include contributions from the scalar field and its coupling with spacetime curvature, while these contributions typically encompass kinetic terms, nonminimal couplings to curvature invariants such as the Ricci scalar, and, in more general scenarios, scalar self interaction potentials.  The resulting field equations are obtained by independently varying the action with respect to the metric tensor and the scalar field, leading to a coupled system of nonlinear differential equations characteristic of scalar tensor theories~\cite{DeFelice2010}. One of the principal motivations for investigating such theories is their ability to address limitations of GR, particularly on cosmological scales, while scalar tensor models offer a natural mechanism for modifying gravitational dynamics and extensively explored as alternatives to the $\Lambda$CDM paradigm~\cite{Clifton2012,DeFelice2010}.
Furthermore, investigations into gravitational collapse and BH formation under modified corrections have revealed important modifications to the classical picture, suggesting singularity avoidance and new phenomenological scenarios for horizon dynamics \cite{bojowald2005singularity,bojowald2009non,munch2021effective,husain2022semi,giesel2021towards,yousaf2024FBH,donmez2026relativistic,donmez2026accretion}, while a notable class within this scenario involves models in which the scalar field couples directly to matter, such couplings proved particularly useful in the study of compact astrophysical systems, including neutron stars and black holes (BH)~\cite{DamourEspositoFarese1993,DamourEspositoFarese1996,Harada1998}. Notably, they can account for the observed accelerated expansion of the universe without invoking a cosmological constant, instead, the scalar field dynamically drives this expansion, effectively mimicking dark energy~\cite{Copeland2006,AmendolaTsujikawa2010}. In this context, the Freund-Nambu model provides an important theoretical laboratory for examining alternative explanations of cosmic acceleration and the nature of dark energy. Over time, the Freund-Nambu theory applied to a variety of physical scenarios, including early universe inflation, large scale cosmological dynamics, BH physics, and gravitational wave phenomena~\cite{Kanti1996,Sotiriou2012}. 

From a mathematical standpoint, singularities are characterized by divergences in curvature invariants. Such as the Kretschmann scalar which signals the breakdown of classical spacetime descriptions, whereas a naked singularity arises when causal geodesics originating arbitrarily close to the singular region. It can extend to infinity without encountering an event horizon. Whereas a well-known example is the Janis-Newman-Winicour  solution~\cite{Janis68}. These type of solutions generalizes the Schwarzschild spacetime by incorporating a massless scalar field, while an additional scalar parameter significantly alters the geometry. The analysis of test particle motion in static, spherically symmetric spacetimes serves as a powerful tool to study and for good examination in gravitational theories. One can extract important physical quantities such as the innermost stable circular orbit (ISCO), specific energy and angular momentum, as well as radial and vertical epicyclic frequencies~\cite{stuchlik2013,bambi2017,Aliev:2006qi,2026AnPhy.48670332N,Turakhonov:2024smp,Turakhonov:2024xfg,Ibrokhimov:2024hxg,Umarov:2025ihy,Umarov:2025wzm,Turimov:2024hwh,Turimov:2024tvt,Turimov:2025odi,Turimov:2025tmf,Turimov:2022iff,Boboqambarova:2021cbf,Turimov:2021jgk,Turimov:2020fme, Gogoi:2026obd}. To establish a meaningful comparison between theoretical predictions and observational data, reliable statistical techniques are required~\cite{sharma2017,foreman-mackey2013,Shabbir:2026qlh,Shermatov:2025ljg,Zahra:2025tdo,Shermatov:2025rpj,Turimov:2024orr}. In the context of BH astrophysics, this analyses of quasi-periodic oscillations data from sources such as XTE J1550-564 and GRS 1915+105 extensively utilized to infer BH properties and to test deviations from standard GR based metrics~\cite{shafee2006,kolos2023,Hoshimov:2025tdx,stuchlik2021,bambi2018}.

In the present study, we focus on Freund-Nambu scalar tensor gravity as a useful scenario for examining the influence of scalar degrees of freedom on compact entity spacetimes and their astrophysical signatures. While in this setting, the scalar field is not merely an auxiliary ingredient, but an active geometric component that modifies the gravitational configuration itself which makes it possible to investigate how scalar induced corrections affect the background geometry. It also investigate particle dynamics, and potentially observable strong field phenomena. In this analysis one can observe that a central aspect is the construction of an exact analytical solution in the vacuum limit of the theory. In this scenario matter sector is neglected and the scalar field is taken to be massless which is still rich enough to capture the essential geometric role of the scalar field. While allowing a transparent examination of the resulting spacetime structure. However in this way, the model provides a natural setting to isolate purely gravitational effects generated by the scalar interaction~\cite{Freund1968PR,Janis68}, however its compatibility with observational data remains under investigation~\cite{Will2014}. In this regard, simplified dark energy models constructed within this scenario also explored, highlighting the role of a time dependent scalar field in shaping late time cosmological behavior~\cite{Dudko2016GC}, while the resulting geometry can be viewed as a scalar field deformed extension of the well-known spacetime which is call Janis-Newman-Winicour. This modification is especially important because it can produce naked singularity-type configurations rather than standard black hole horizons, thereby offering a useful theoretical arena for studying deviations from GR in strong gravitational fields. As a consequence, the scalar sector influences not only the spacetime geometry itself but also the way matter moves within it, making the analysis of particle trajectories especially significant for identifying observational signatures of the theory.

The study of the effective potential reveals that the scalar parameters substantially reshape the orbital landscape of the system, while stable and unstable circular orbits are shifted relative to the corresponding Schwarzschild behavior. One also observes that the depth as well as the location of the orbital potential become highly sensitive to the scalar field contributions, while these changes directly affect the structure of strong field motion and demonstrate that even modest scalar modifications may leave a clear imprint on the dynamics of matter near compact emtities. Particular attention is given to the ISCO and the marginally bound orbit, since both play a key role in accretion physics and strong-gravity phenomenology. While our analysis indicates that the scalar interaction can either pull the innermost stable orbit inward or shift it outward, depending on the sign and strength of the coupling, whereas such type of asymmetry highlights the fact that scalar effects may either enhance or weaken orbital stability~\cite{Aliev:2006qi,stuchlik2013,kolos2023}. Beyond circular motion, this scenario also admits a family of bound periodic trajectories displaying the characteristic zoom whirl behavior known from strong field gravity how particles repeatedly move from distant regions into the deep gravitational well before returning outward again. These trajectories provides study the orbital morphology of the Freund-Nambu spacetime~\cite{Levin:2008mq,Glampedakis:2002ya}. An important outcome of the orbital analysis is that the transition toward the separatrix regime is strongly affected by the scalar parameters, however, as a particle approaches this regime, the number of rapid revolutions near the central entity increases, signaling an enhancement of relativistic precession and strong field trapping effects. This behavior shows that the scalar sector can significantly alter the fine structure of bound motion, with direct consequences for the timing properties and topology of orbits in naked singularity backgrounds~\cite{Levin:2008mq,Glampedakis:2002ya}, whereas the astrophysical significance of these results becomes even more pronounced when gravitational wave emission is considered. In the case of an extreme mass ratio inspiral, where a stellar mass compact entity moves around a supermassive central source, the orbital properties of the background geometry are directly encoded in the emitted waveform. Since such systems are among the prime targets of future space based detectors, the Freund-Nambu spacetime offers a timely and relevant setting for investigating how scalar tensor effects may appear in gravitational wave observations~\cite{BabakEtAl:2007,Hughes:2000ssa,Thorne:1980ru,Meng:2024cnq,Zhao:2024exh}. A particularly interesting result of the present work is that variations in the scalar parameters can generate pronounced dephasing in the gravitational wave signal, even when the overall spatial shape of the orbit remains nearly unchanged, whereas this shows that waveform timing can serve as a sensitive probe of the underlying scalar structure of the spacetime. Therefore, the combination of orbital dynamics, zoom whirl behavior, and gravitational wave modeling provides a promising route for testing Freund-Nambu scalar tensor gravity and for constraining its coupling parameters in the strong field regime~\cite{BabakEtAl:2007,Glampedakis:2002ya,Meng:2024cnq,Zhao:2024exh}.

This paper is structured as follows: In Section \ref{Sec:Formulations}, we present the mathematical formulation of the Freund-Nambu scalar-tensor gravity and derive the exact spherically symmetric vacuum solution. Section \ref{sec.4} explores test particle dynamics and the effective potential under the influence of direct scalar coupling. Section \ref{Sec.5} analyzes the properties of the marginally bound and innermost stable circular orbits. In Section \ref{sec.6}, we classify periodic orbits and investigate their zoom-whirl behavior, cataloging them by their topological parameters. Section \ref{sec.7} computes the gravitational wave radiation emitted by extreme mass-ratio inspirals, highlighting the macroscopic dephasing induced by scalar parameters. Finally, we summarize our main conclusions in Section \ref{Sec:Conclusions}.

\section{Freund-Nambu scalar-tensor gravity\label{Sec:Formulations}}

In the context of Freund–Nambu scalar–tensor gravity, the action describing the Einstein–scalar field system, formulated in geometrized units with $G = c = 1$, is written as~\cite{Freund1968PR}
\begin{align}\label{action}
{\cal S} = \frac{1}{16\pi} \int d^4x, \sqrt{-g} \left(R - \frac{2\partial_i\varphi\partial^i\varphi}{1 + 2q\varphi} + 2\mu^2 \varphi^2 \right) + {\cal S}_M \ ,
\end{align}
where $g = |g_{ij}|$ denotes the determinant of the spacetime metric, $R$ is the Ricci scalar, and $\varphi$ is a real scalar field of mass $\mu$. The parameter $q$ controls the strength of the coupling between the scalar field and the spacetime geometry. The term ${\cal S}_M$ denotes the matter action, built from the matter fields $Q_M$ and minimally coupled to the conformally rescaled metric $(1 + 2q\varphi)g_{ij}$, so that ${\cal S}_M = {\cal S}_M\big[(1 + 2q\varphi)g_{ij},Q_M\big]$. By varying the total action \eqref{action} with respect to the metric $g_{ij}$ and the scalar field $\varphi$, we obtain the field equations that describe the dynamics of the coupled Einstein–scalar system, namely

\begin{align}\label{eq}
&G_{ij}=R_{ij}-\frac{1}{2}g_{ij}R=T_{ij}+T^M_{ij} \ , \\
\label{eqf}
&(\square - \mu^2)\varphi = q\left( T + T^M \right) \ ,
\end{align}

where $G_{ij}$ denotes the Einstein tensor and $\square = -\nabla^i\nabla_i$ represents the covariant d’Alembert operator. The energy–momentum tensor $T_{ij}$ associated with the scalar field is given by

\begin{align}
T_{ij} = \frac{2\partial_i\varphi\partial_j\varphi}{1 + 2q\varphi} - g_{ij} \left( \frac{\partial_\gamma\varphi,\partial^\gamma\varphi}{1 + 2q\varphi} - \mu^2 \varphi^2 \right) \ ,
\end{align}
where trace is given by the following relation:
\begin{align}
T = g^{ij} T_{ij} = -\frac{2\partial_i\varphi\partial^i\varphi}{1 + 2q\varphi} + 4\mu^2 \varphi^2 \ .
\end{align}

In this work, we seek to obtain an exact analytical solution of the Einstein equations coupled to a massless scalar field within Freund–Nambu theory. To streamline the analysis, we focus on the vacuum configuration by discarding the scalar field mass term (i.e., taking $\mu = 0$) and omitting the matter sector (i.e., setting ${\cal S}_M = 0$). This truncation enables us to concentrate solely on the geometric impact of a self-interacting scalar field in a curved spacetime background. In the next section, we present a detailed derivation of the resulting field equations and examine the associated exact solutions. Upon neglecting the matter term, the Einstein–massless scalar field equations in the Freund–Nambu model can be written as:
\begin{align}
\square\varphi=-\frac{q}{1+2q\varphi}\partial_i\varphi\partial^i\varphi\ .
\end{align}
Accordingly, the equation for the massless scalar field may be expressed in the following form:
\begin{align}\label{KG}
&\frac{1}{\sqrt{-g}}\partial_\mu\left(\sqrt{-g}\partial^\mu\varphi\right)=\frac{q}{1+2q\varphi}\partial_i\varphi\partial^i\varphi\ .
\end{align}
Assume that the scalar field is time-independent and depends solely on the radial coordinate, i.e., $\varphi=\varphi(r)$. Then equation (\ref{KG}) becomes
\begin{align}
&\frac{1}{\sqrt{-g}}\frac{d}{dr}
\left(\sqrt{-g}g^{rr}\frac{d\varphi}{dr}\right)=\frac{q}{1+2q\varphi}g^{rr}\left(\frac{d\varphi}{dr}\right)^2\ .
\end{align}
Hereafter, by introducing the new field \(F = 1 + 2q\varphi\), the equation simplifies to
\begin{align}
\left(\sqrt{-g}g^{rr}\frac{dF}{dr}\right)^{-1}\frac{d}{dr}
\left(\sqrt{-g}g^{rr}\frac{dF}{dr}\right)=\frac{1}{2F}\left(\frac{dF}{dr}\right)\ ,
\end{align}
which further can be reads as
\begin{align}
&\frac{d}{dr}\ln\left(\sqrt{-g}g^{rr}\frac{dF}{dr}\right)=\frac{d}{dr}\ln\sqrt{F}\ ,
\end{align}
Upon integration of the preceding equation, one obtains
\begin{align}
&\frac{dF}{\sqrt{F}}=\frac{2C_1dr}{\sqrt{-g}g^{rr}}\ ,
\end{align}
or
\begin{align}\label{solF}
&\sqrt{F}=\int\frac{C_1dr}{\sqrt{-g}g^{rr}}+C_2\ ,
\end{align}
where \(C_1\) and \(C_2\) denote integration constants. Once the spacetime metric has been determined, the configuration of the scalar field can be obtained directly from expression \eqref{solF}. The physical interpretation of these constants will be discussed at a later stage. Consequently, the spherically symmetric solution in the Freund–Nambu theory \cite{Davlataliev:2026vkx} can be written in the form
\begin{align}\label{JNW}
ds^2&=-\left(1-\frac{2M}{nr}\right)^{n}dt^2+\left(1-\frac{2M}{nr}\right)^{-n}dr^2+r^2\left(1-\frac{2M}{nr}\right)^{1-n}(d\theta^2+\sin^2\theta d\phi^2)\ ,
\end{align}
where \(M\) denotes the mass of the object, and the parameter \(n\) arises due to the presence of the scalar field. The associated scalar field is given by
\begin{align}
\varphi=\frac{\sqrt{1-n^2}}{2}\ln\left(1-\frac{2M}{nr}\right)\left[1+\frac{q}{2}\frac{\sqrt{1-n^2}}{2}\ln\left(1-\frac{2M}{nr}\right)\right]\ .    
\end{align}
In this particular case, the metric (\ref{JNW}) reduces to the well-known Janis–Newman–Winicour solution~\cite{Janis68}. Thus, the spacetime geometry is described by equation~\eqref{JNW}, while the corresponding scalar field takes the form
\begin{align}
\varphi=\frac{\sqrt{1-n^2}}{2}\ln\left(1-\frac{2M}{nr}\right)\ .    
\end{align}

\section{Particle Dynamics and Effective Potential \label{sec.4}}

To understand the astrophysical implications of the Freund-Nambu scalar-tensor solution \cite{Freund1968PR}, we investigate the motion of test particles in the background of the Janis-Newman-Winicour (JNW) spacetime \cite{Janis68}. In this framework, a direct linear coupling between the test particle and the scalar field is introduced via an additional parameter $g_s$. Consequently, the effective mass of the test particle becomes $m_* = m(1+g_s\varphi)$. The dynamics of a massive test particle are governed by the modified Lagrangian:
\begin{equation}
\overline{L} = \frac{1}{2}(1+g_s\varphi)g_{\mu\nu}u^\mu u^\nu
\end{equation}

Due to the spherical symmetry of the metric, we can restrict the motion of the test particle to the equatorial plane without any loss of generality, setting $\theta = \pi/2$ and $\dot{\theta} = 0$. 

The spacetime admits two Killing vectors, $\partial/\partial t$ and $\partial/\partial \phi$, which correspond to the conserved specific energy $E$ and specific angular momentum $L$ of the particle. Due to the scalar interaction, these constants of motion are modified to:
\begin{align}
E &= -(1+g_s\varphi)g_{tt}u^t = (1+g_s\varphi)\left(1-\frac{2M}{nr}\right)^n \dot{t}, \label{energy} \\
L &= (1+g_s\varphi)g_{\phi\phi}u^\phi = (1+g_s\varphi)r^2\left(1-\frac{2M}{nr}\right)^{1-n} \dot{\phi}. \label{angmom}
\end{align}

Using the normalization condition for the four-velocity of the particle, the radial equation of motion takes the form:
\begin{equation}
(1+g_s\varphi)^2\dot{r}^2 = E^2 - V(r)
\label{radial_eq}
\end{equation}
where the effective potential $V(r)$ that governs the radial motion is defined as:
\begin{equation}
V(r) = \left(1-\frac{2M}{nr}\right)^n (1+g_s\varphi)^2 + \frac{L^2}{r^2}\left(1-\frac{2M}{nr}\right)^{2n-1}
\label{V_eff}
\end{equation}

\begin{figure}[htbp]
\centering
\includegraphics[width=0.8\textwidth]{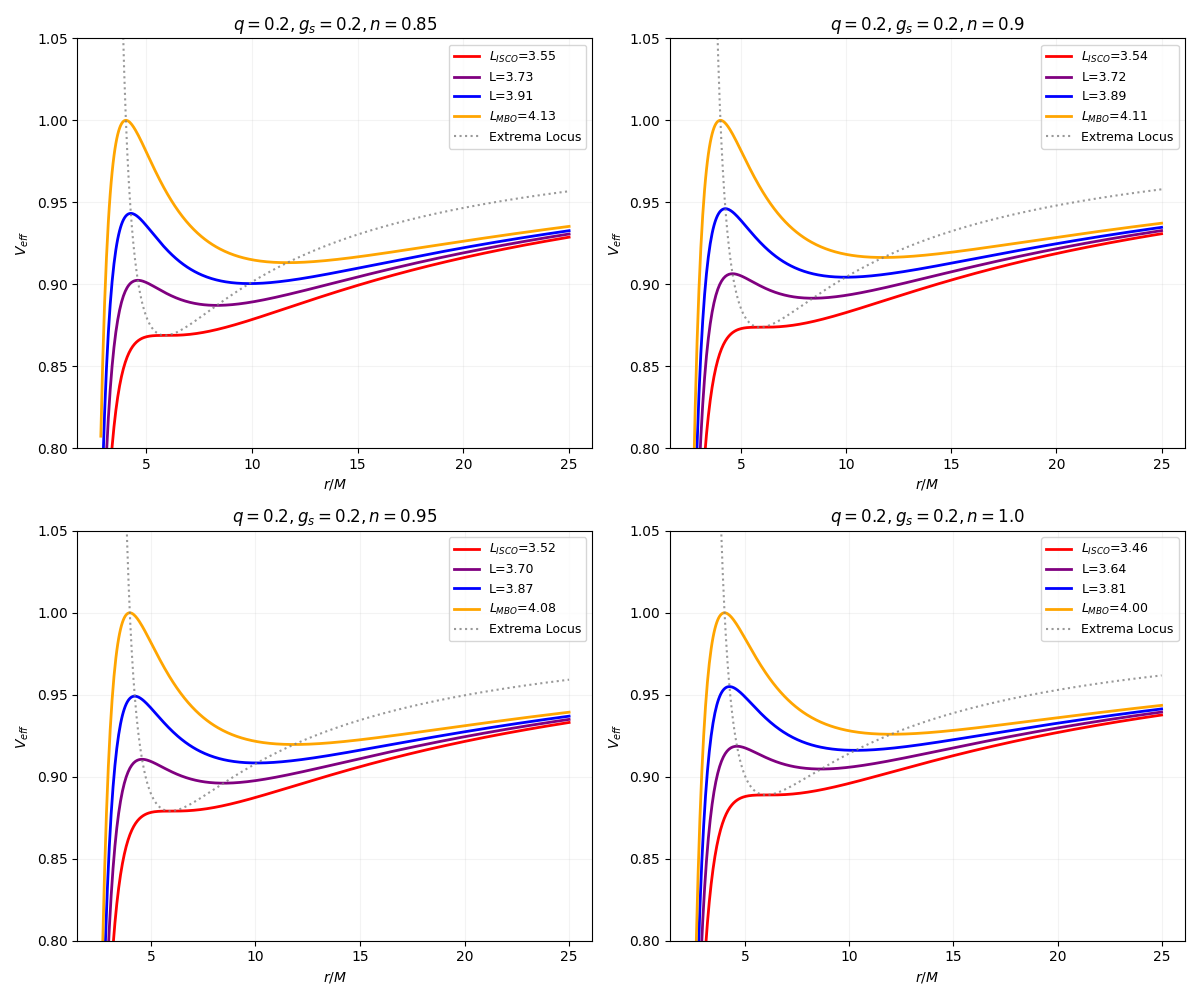}
\caption{The effective potential $V(r)$ for a test particle in the Freund-Nambu JNW spacetime for different values of scalar parameter $n$, with fixed scalar couplings $q$ and $g_s$. The extrema represent stable and unstable circular orbits.}
\label{fig:potential}
\end{figure}

\begin{figure}[htbp]
\centering
\includegraphics[width=0.8\textwidth]{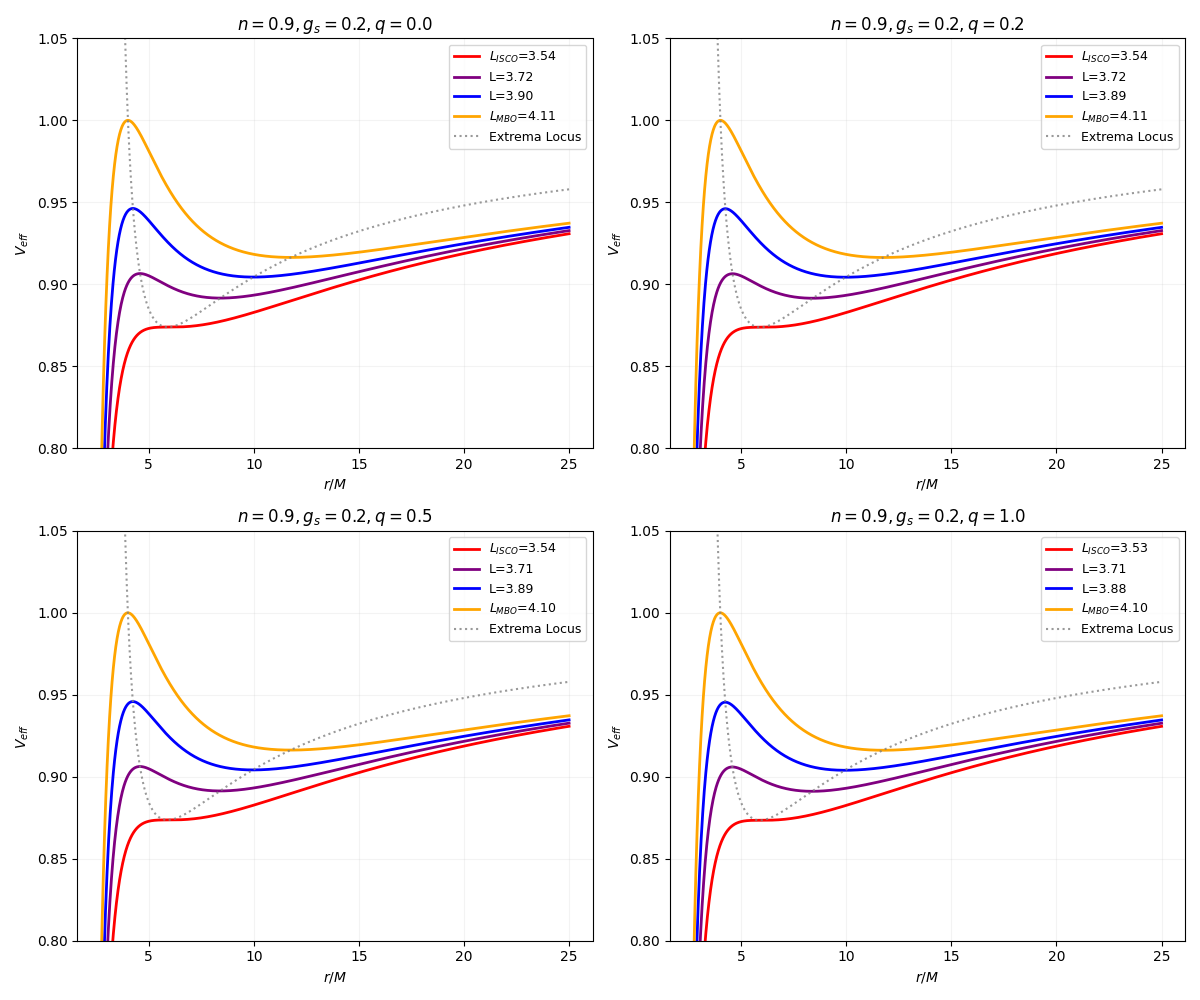}
\caption{The effective potential $V(r)$ for a test particle in the Freund-Nambu JNW spacetime for different values of parameter $q$. The extrema represent stable and unstable circular orbits.}
\label{fig:potential2}
\end{figure}

\begin{figure}[htbp]
\centering
\includegraphics[width=0.8\textwidth]{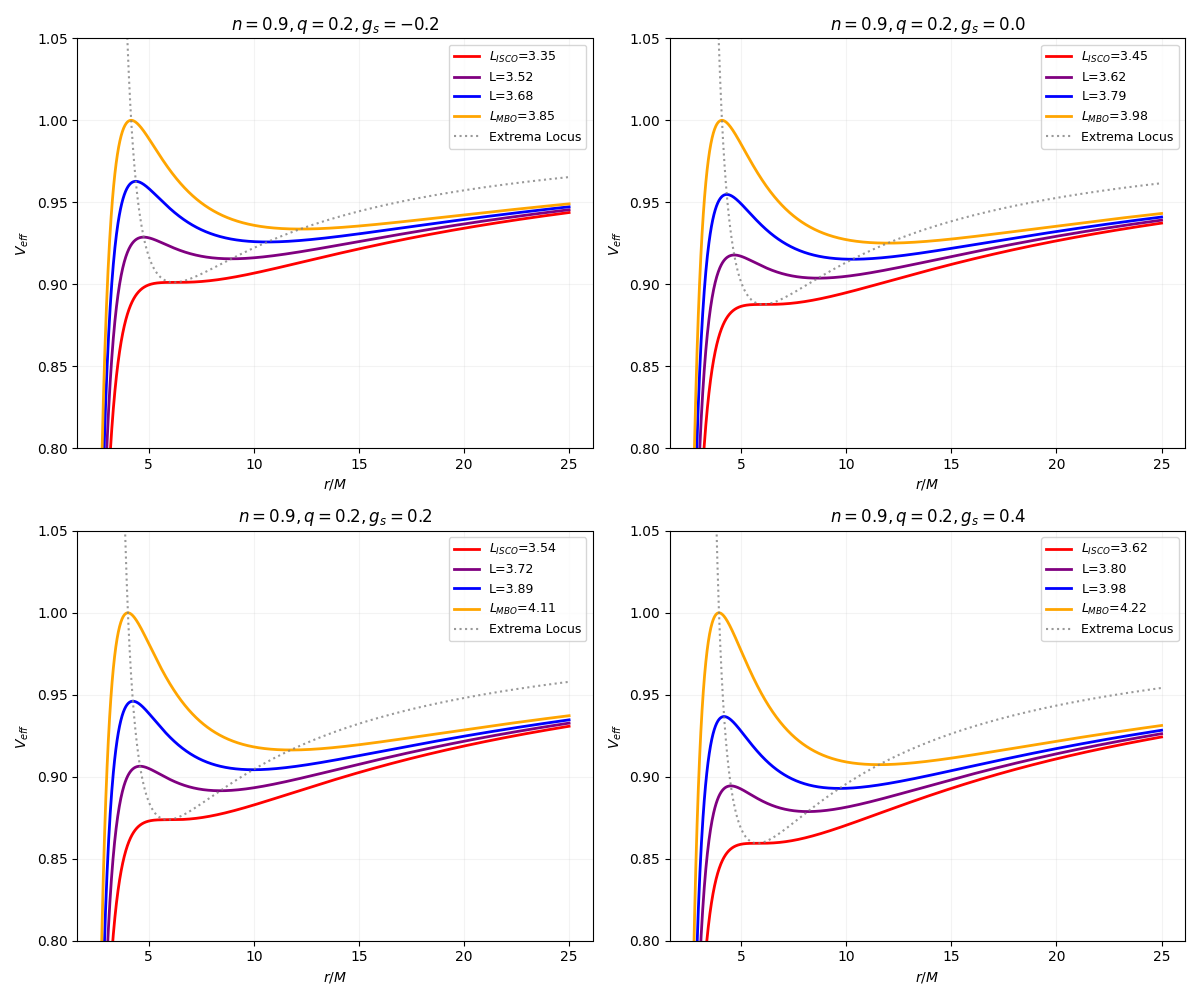}
\caption{The effective potential $V(r)$ for a test particle in the Freund-Nambu JNW spacetime for different values of parameter $g_s$. The extrema represent stable and unstable circular orbits.}
\label{fig:potential3}
\end{figure}

The shape of the effective potential completely determines the regions of allowed motion for test particles. In Fig.~\ref{fig:potential}, we plot the effective potential for various values of the scalar parameter $n$ while keeping the scalar couplings fixed. To explicitly demonstrate the influence of the scalar interactions, we further plot the effective potential for varying values of the geometric coupling $q$ in Fig.~\ref{fig:potential2}, and for varying values of the scalar-particle coupling $g_s$ in Fig.~\ref{fig:potential3}. The presence of the scalar field alters the depth and the location of the minimum (corresponding to stable circular orbits) and the maximum (unstable circular orbits) of the potential compared to the standard Schwarzschild case \cite{stuchlik2013, bambi2017}. As these parameters deviate from unity or zero, the potential barrier shifts, significantly modifying the strong-field orbital dynamics.

\section{Marginally Bound and Innermost Stable Circular Orbits \label{Sec.5}}

The properties of circular orbits are fundamental for modeling accretion disks around compact objects \cite{Aliev:2006qi}. Circular orbits are determined by the conditions $\dot{r} = 0$ and $\partial_r V(r) = 0$, which fix the conserved energy and angular momentum of the particle at a given orbital radius. The Innermost Stable Circular Orbit (ISCO) represents the smallest radius at which a test particle can maintain a stable circular orbit. It is determined by the inflection point of the effective potential, satisfying the simultaneous conditions:
\begin{equation}
V(r_{\rm ISCO}) = E^2, \quad \frac{dV}{dr}\bigg|_{r=r_{\rm ISCO}} = 0, \quad \frac{d^2V}{dr^2}\bigg|_{r=r_{\rm ISCO}} = 0
\label{circular_cond}
\end{equation}

\begin{figure}[htbp]
\centering
\includegraphics[width=0.32\textwidth]{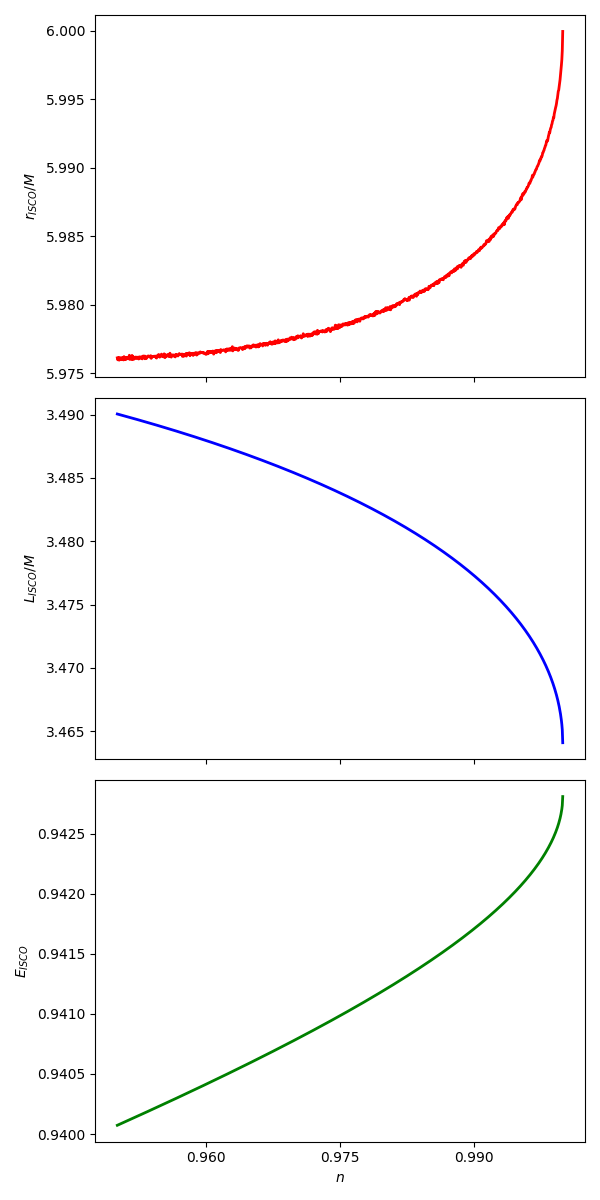}
\includegraphics[width=0.32\textwidth]{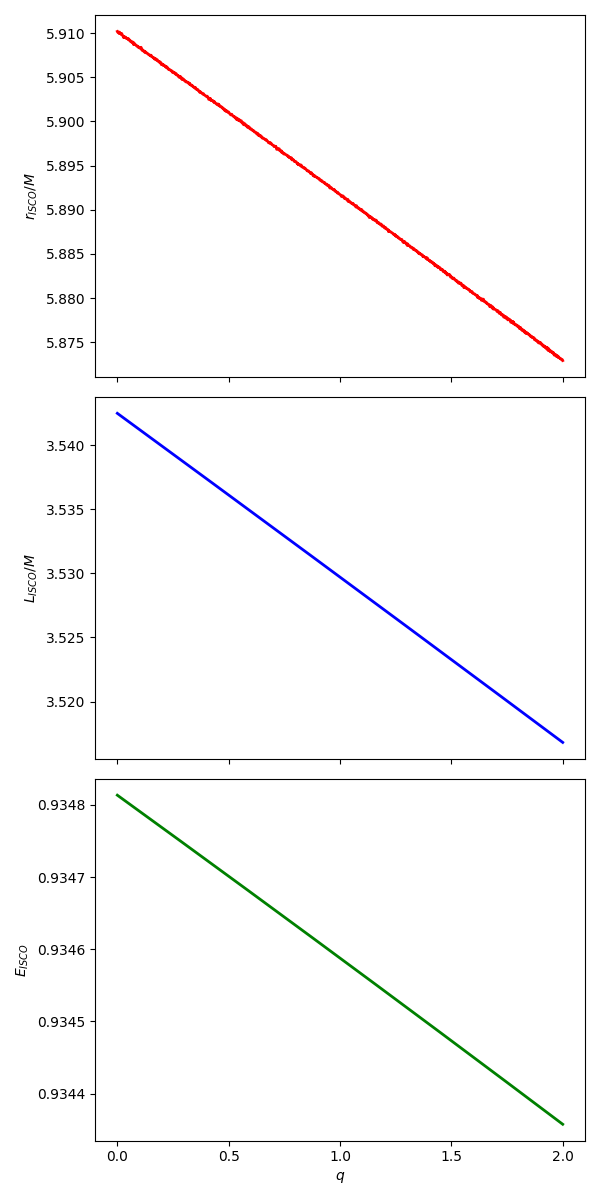}
\includegraphics[width=0.32\textwidth]{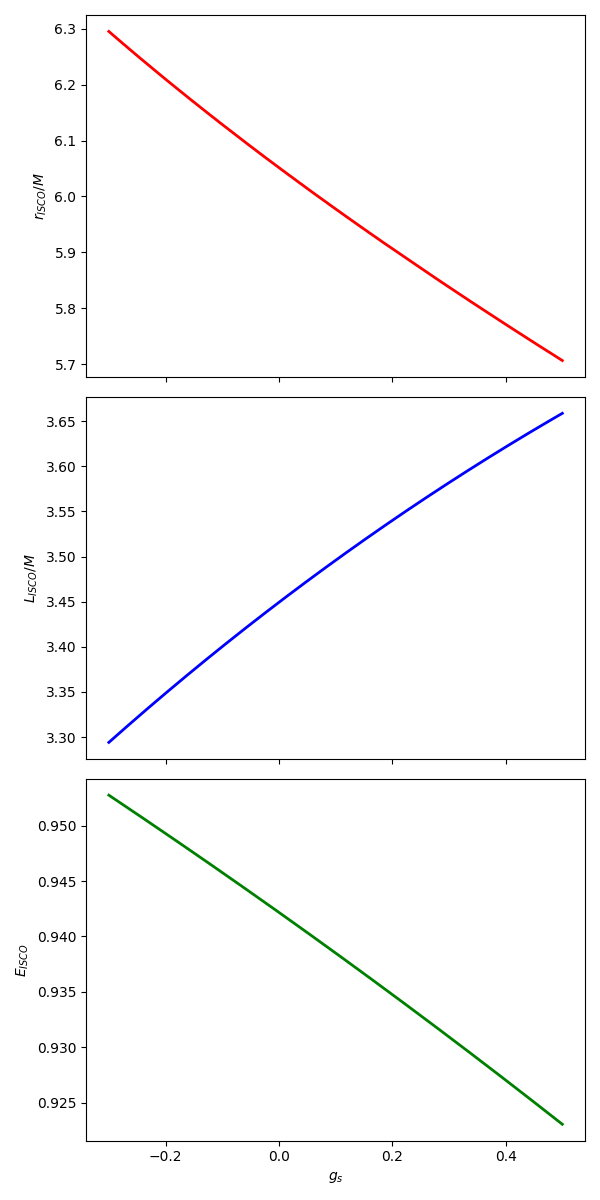}
\caption{Dependence of the ISCO radius on the model parameters. The fixed values, wherever applicable, are: $n=0.9$, $q=0.2$, and $g_s=0.2$.}
\label{fig:isco}
\end{figure}

In Fig.~\ref{fig:isco}, we present the variation of the ISCO radius, specific angular momentum, and specific energy with respect to the spacetime parameter $n$, the geometric coupling $q$, and the particle coupling $g_s$. The presence of the scalar field modifies both the location and stability properties of the circular orbits. Our numerical results show a clear asymmetry in the ISCO behavior depending on the sign of the particle coupling constant. In particular, the ISCO radius decreases for $g_s > 0$, indicating that attractive scalar interactions enhance orbital stability at smaller radii, whereas for $g_s < 0$ the ISCO shifts outward, signaling a destabilizing effect.

\begin{figure}[htbp]
\centering
\includegraphics[width=0.45\textwidth]{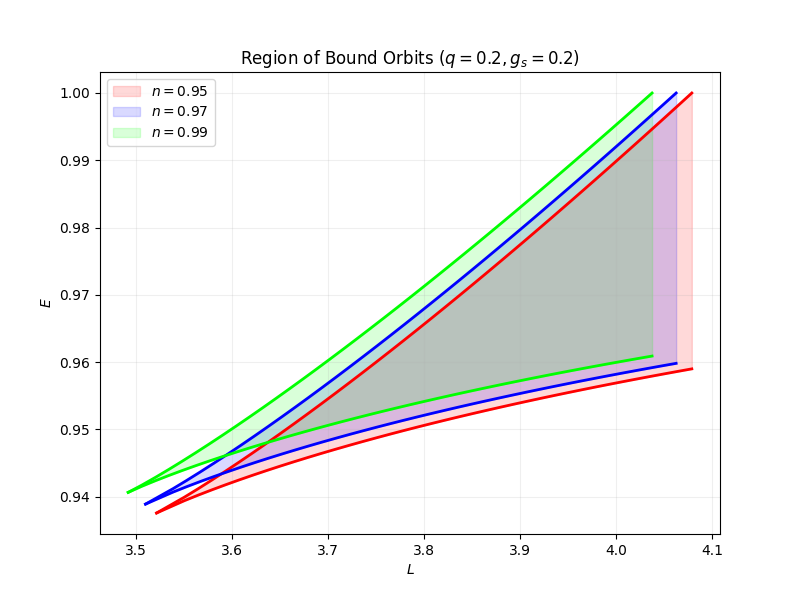}
\includegraphics[width=0.45\textwidth]{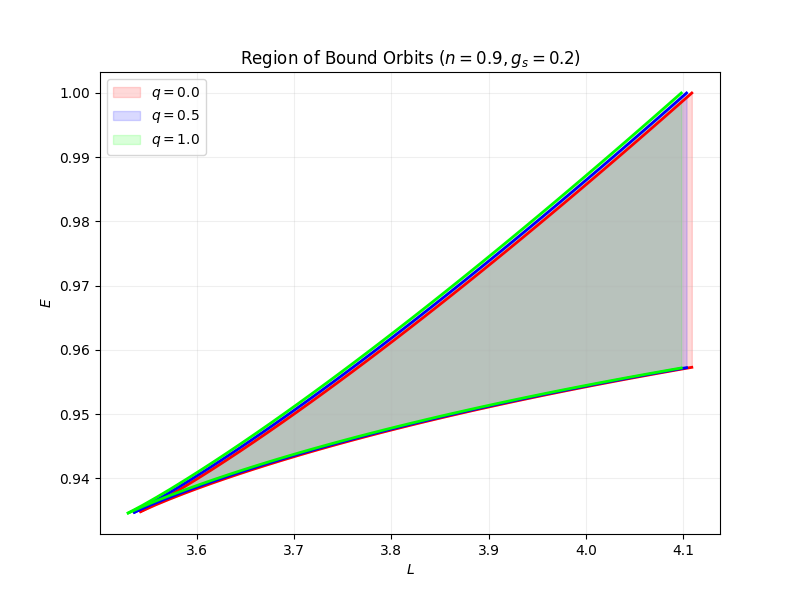}
\includegraphics[width=0.45\textwidth]{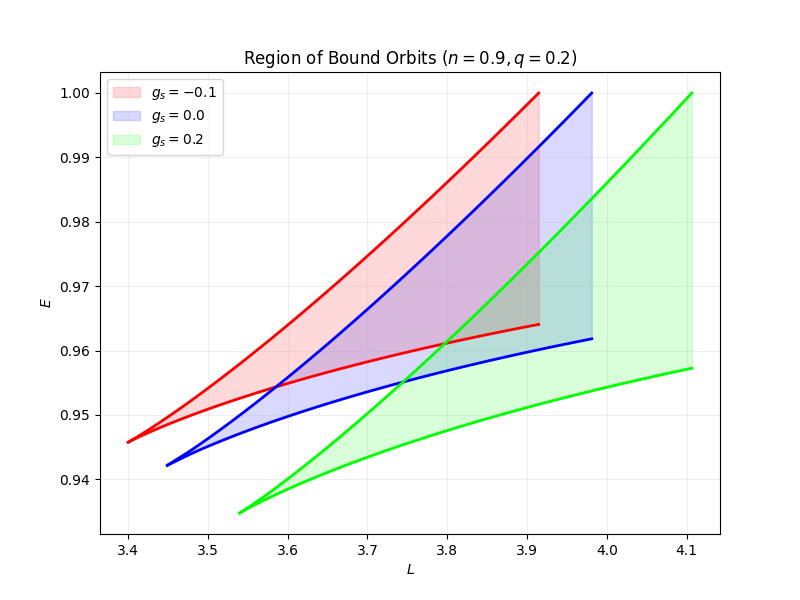}
\caption{The parameter space $(E, L)$ allowing for bound orbits.}
\label{fig:paramspace}
\end{figure}

Similarly, the Marginally Bound Orbit (MBO), defined by the condition $E=1$, dictates the threshold between bound and unbound particle states. Fig.~\ref{fig:paramspace} illustrates the allowed parameter space ($E$ versus $L$) for bound orbits. The shaded regions delineate the combinations of specific energy and angular momentum for which bounded oscillatory motion is physically permitted. The introduction of the scalar field distorts this parameter space relative to the Schwarzschild geometry, highlighting how the non-trivial scalar profile and particle coupling influence the phase space of bound trajectories.

\section{Periodic Orbits and Zoom-Whirl Behavior \label{sec.6}}

Among the bound trajectories, periodic orbits are of particular interest as they manifest the intricate zoom-whirl behavior characteristic of strong-field gravity \cite{Levin:2008mq, Glampedakis:2002ya}. An orbit is periodic if the ratio of the azimuthal frequency to the radial frequency is a rational number. We characterize this topology using the rational parameter $q_{\rm rat}$, defined as:
\begin{equation}
q_{\rm rat} = \frac{\Delta \phi}{2\pi} - 1 = w + \frac{v}{z},
\label{q_param}
\end{equation}
where $\Delta \phi$ is the total accumulated azimuthal angle during one complete radial oscillation from the periapsis $r_p$ to the apoapsis $r_a$ and back. The integers $z$, $w$, and $v$ denote the zoom number, whirl number, and vertex number, respectively. Using the modified equations of motion, the azimuthal shift $\Delta \phi$ is given by:
\begin{equation}
\Delta \phi = 2 \int_{r_p}^{r_a} \frac{\dot{\phi}}{\dot{r}} dr = 2 \int_{r_p}^{r_a} \frac{L}{r^2 \left(1-\frac{2M}{nr}\right)^{1-n} \sqrt{E^2 - V(r)}} dr.
\end{equation}

\begin{figure}[htbp]
\centering
\includegraphics[width=0.9\textwidth]{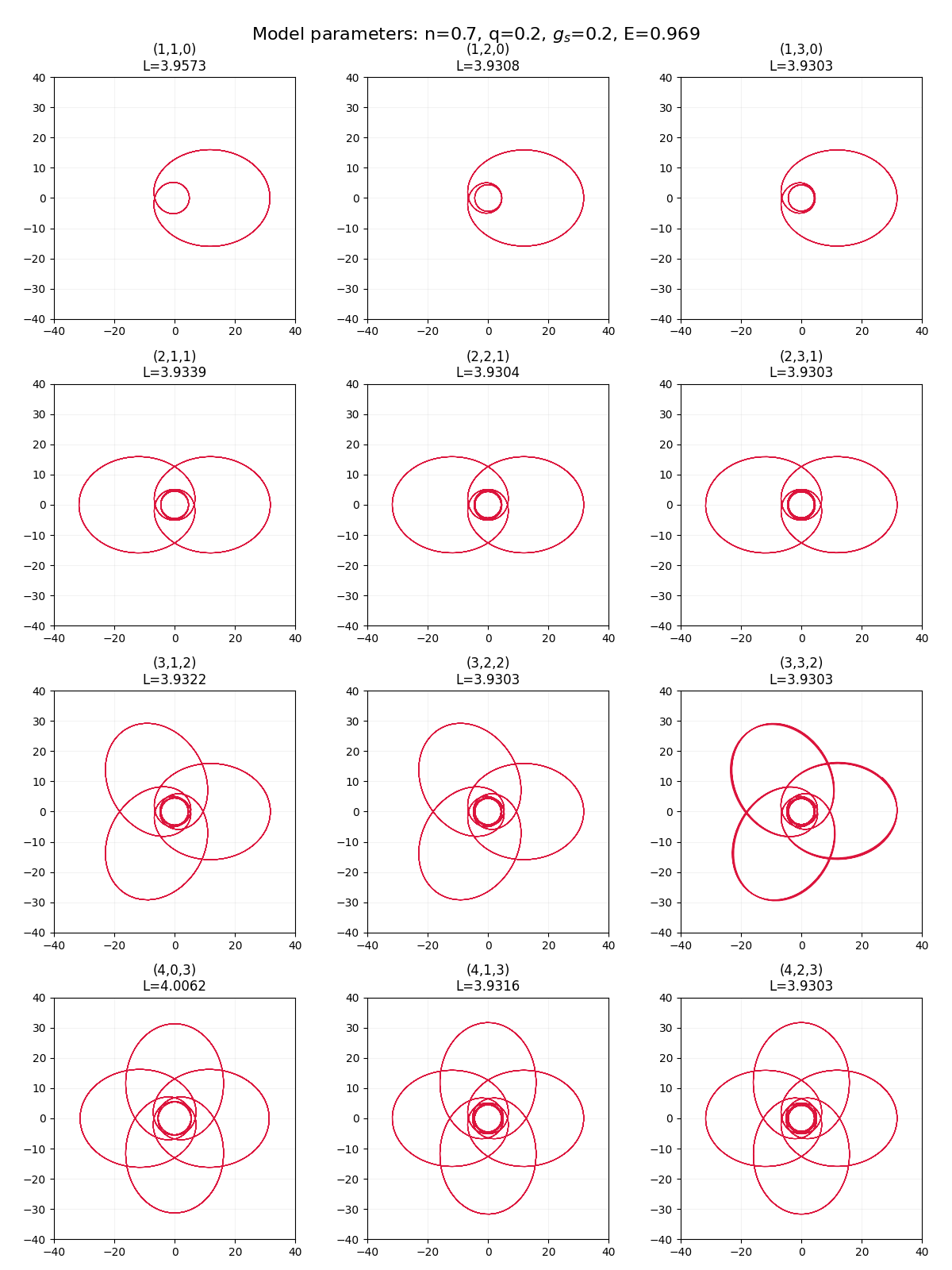}
\caption{Taxonomy of periodic orbits categorized by the $(z, w, v)$ topology for a fixed specific energy $E = 0.969$. By incrementally varying the specific angular momentum $L$, the trajectories display an increasing number of whirls and petals.}
\label{fig:wz01}
\end{figure}

\begin{figure}[htbp]
\centering
\includegraphics[width=0.9\textwidth]{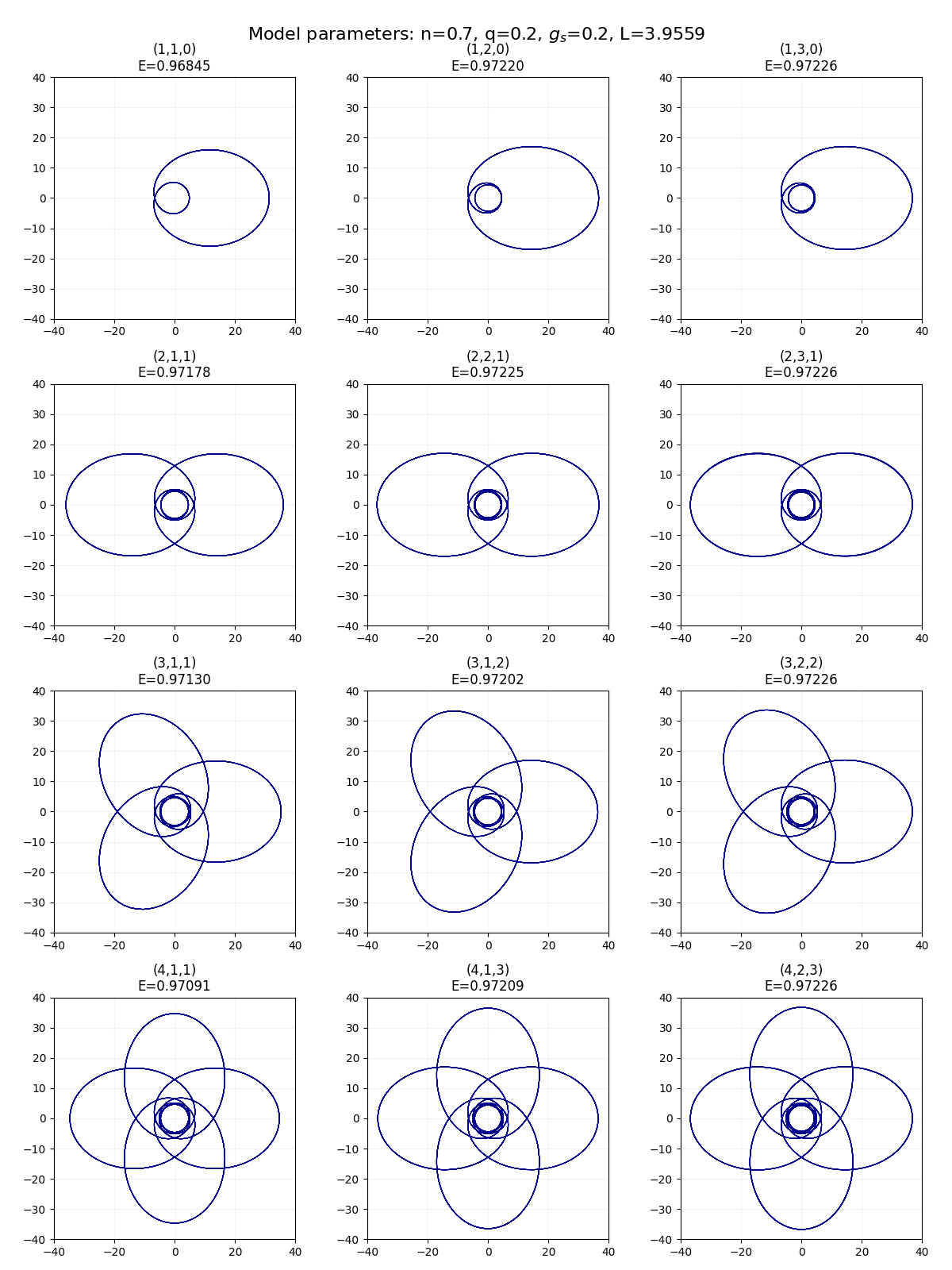}
\caption{Taxonomy of periodic orbits categorized by the $(z, w, v)$ topology for a fixed specific angular momentum $L = 3.9559$. Varying the specific energy $E$ reveals deep-field periapsis precession dynamics.}
\label{fig:wz02}
\end{figure}

By tracking the integers $(z, w, v)$, we can map the rich morphological diversity of these trajectories. In Fig.~\ref{fig:wz01} and Fig.~\ref{fig:wz02}, we present a systematic taxonomy of these periodic orbits in the Freund-Nambu spacetime. Fig.~\ref{fig:wz01} demonstrates the evolution of orbit topologies as the angular momentum $L$ is tuned for a fixed specific energy, while Fig.~\ref{fig:wz02} displays the inverse relationship by tuning the energy $E$ for a fixed angular momentum.

\begin{figure}[htbp]
\centering
\includegraphics[width=0.6\textwidth]{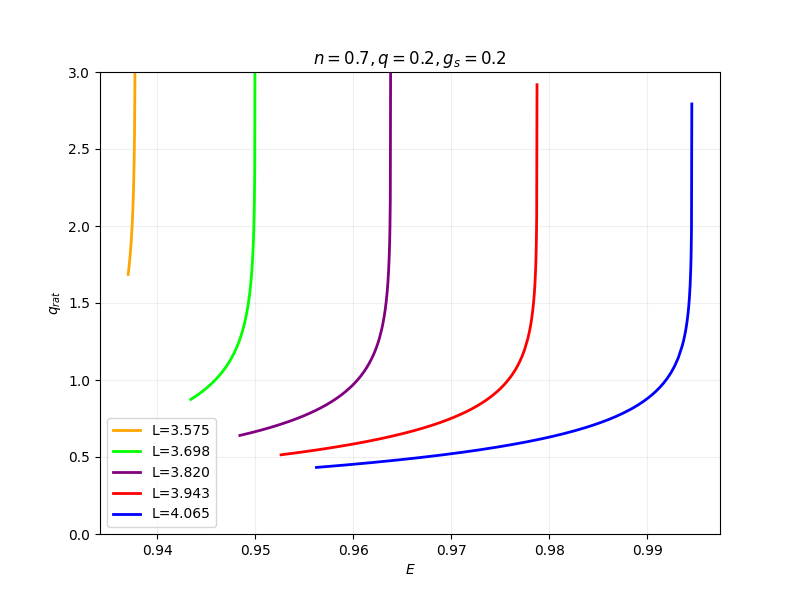}
\caption{The rational parameter $q_{\rm rat}$ plotted as a function of the specific energy $E$ for fixed values of angular momentum $L$. The rapid growth corresponds to the onset of zoom-whirl dynamics near the separatrix.}
\label{fig:qvsE}
\end{figure}

\begin{figure}[htbp]
\centering
\includegraphics[width=0.8\textwidth]{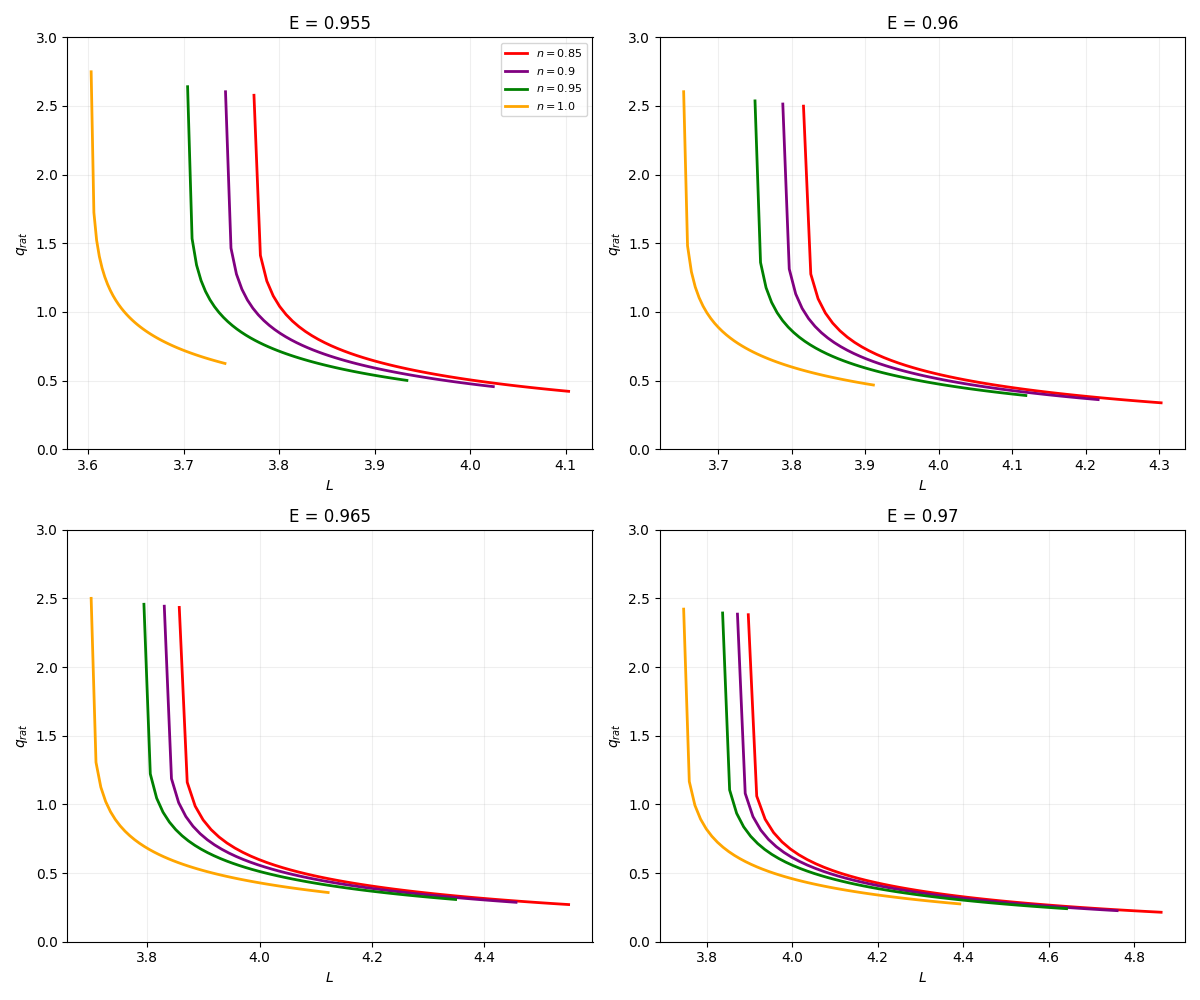}
\caption{The rational parameter $q_{\rm rat}$ plotted against the specific angular momentum $L$ for fixed energy $E$ for different values of the model parameter $n$. Smaller angular momenta lead to intense periapsis precession and higher $q_{\rm rat}$ values.}
\label{fig:qvsL}
\end{figure}

\begin{figure}[htbp]
\centering
\includegraphics[width=0.8\textwidth]{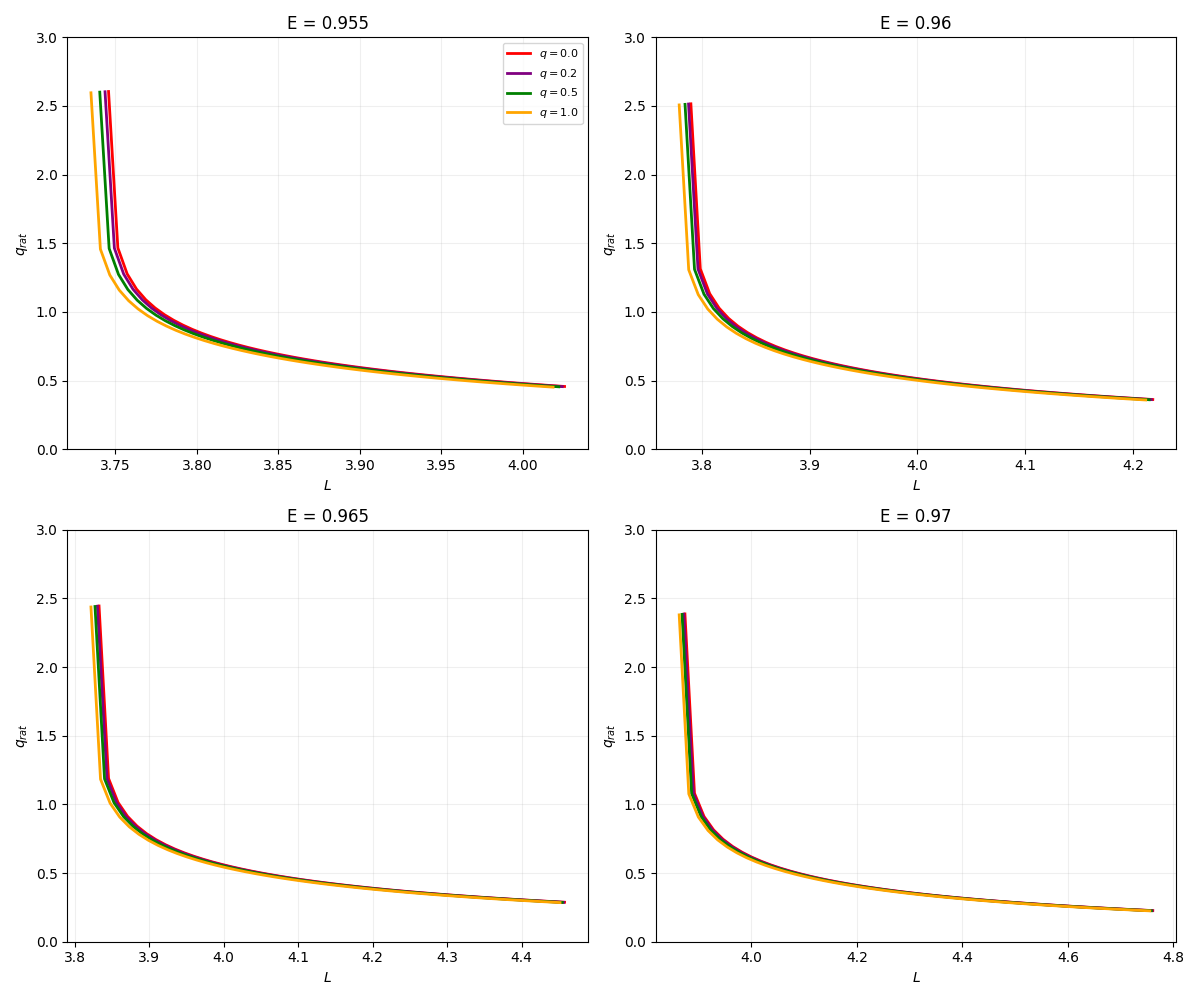}
\caption{The rational parameter $q_{\rm rat}$ plotted against the specific angular momentum $L$ for fixed energy $E$ for different values of the model parameter $q$.}
\label{fig:qvsL2}
\end{figure}

\begin{figure}[htbp]
\centering
\includegraphics[width=0.8\textwidth]{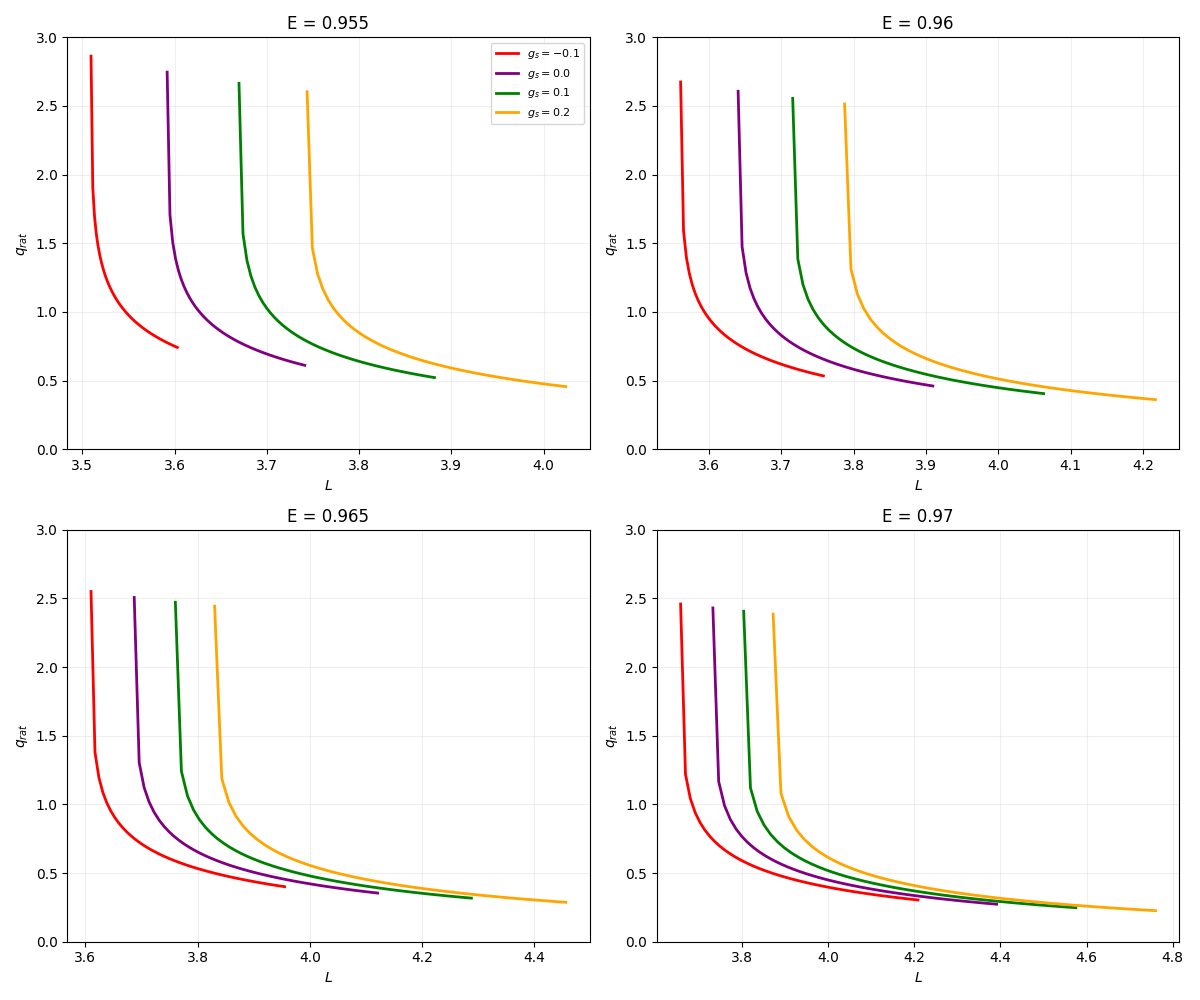}
\caption{The rational parameter $q_{\rm rat}$ plotted against the specific angular momentum $L$ for fixed energy $E$ for different values of the model parameter $g_s$.}
\label{fig:qvsL3}
\end{figure}

The continuous evolution of the rational parameter $q_{\rm rat}$ provides further insight into the separatrix boundary. Fig.~\ref{fig:qvsE} depicts the evolution of $q_{\rm rat}$ as a function of the orbital energy $E$ for fixed values of $L$. As the energy approaches the peak of the effective potential, the particle executes an increasing number of whirls near the periapsis before zooming out to the apoapsis, causing $q_{\rm rat}$ to diverge. Conversely, we demonstrate the behavior of $q_{\rm rat}$ with respect to the angular momentum $L$ for fixed energy states across different parameter regimes. Fig.~\ref{fig:qvsL} reveals this strong-field periapsis precession for different values of the spacetime parameter $n$. Similarly, Fig.~\ref{fig:qvsL2} and Fig.~\ref{fig:qvsL3} illustrate how the non-linear coupling $q$ and the particle coupling $g_s$, respectively, modulate the onset of these zoom-whirl dynamics as the angular momentum decreases.

\section{Gravitational Wave Radiation from Periodic Orbits \label{sec.7}}

To uncover the observable astrophysical signatures of the Freund-Nambu scalar-tensor gravity, we investigate the gravitational wave (GW) emission from a stellar-mass compact object orbiting a supermassive JNW naked singularity. This binary configuration, known as an Extreme Mass-Ratio Inspiral (EMRI), is a primary target for upcoming space-based gravitational wave observatories such as LISA \cite{BabakEtAl:2007}. 

Because the mass of the secondary object ($m$) is extremely small compared to the central mass ($M$), its backreaction on the background spacetime can be safely neglected over short timescales. By applying the adiabatic approximation \cite{Hughes:2000ssa, Glampedakis:2002ya}, the specific energy $E$ and angular momentum $L$ of the particle are considered effectively constant over a single orbital period. To compute the emitted gravitational wave strains, we utilize the Numerical Kludge (NK) method \cite{BabakEtAl:2007}. First, we numerically integrate the exact modified geodesic equations in the curved Freund-Nambu spacetime to obtain the spatial coordinates $(r, \theta, \phi)$ as functions of proper time. These coordinates are mapped to a flat pseudo-Minkowski Cartesian basis:
\begin{equation}
x = r\sin\theta\cos\phi, \quad y = r\sin\theta\sin\phi, \quad z = r\cos\theta
\end{equation}

Using the slow-motion, weak-field quadrupole approximation \cite{Thorne:1980ru}, the spatial strain tensor $h_{ij}$ is evaluated based on the particle's position, velocity, and acceleration:
\begin{equation}
h_{ij} = \frac{2m}{D_L}\left(a_i x_j + a_j x_i + 2 v_i v_j \right)
\end{equation}
where $D_L$ represents the luminosity distance. Finally, by projecting this tensor onto the observer's transverse-traceless plane, we extract the observable plus ($h_+$) and cross ($h_\times$) polarizations \cite{Meng:2024cnq, Zhao:2024exh}.

\begin{figure*}[htbp]
\centering
\includegraphics[width=0.9\textwidth]{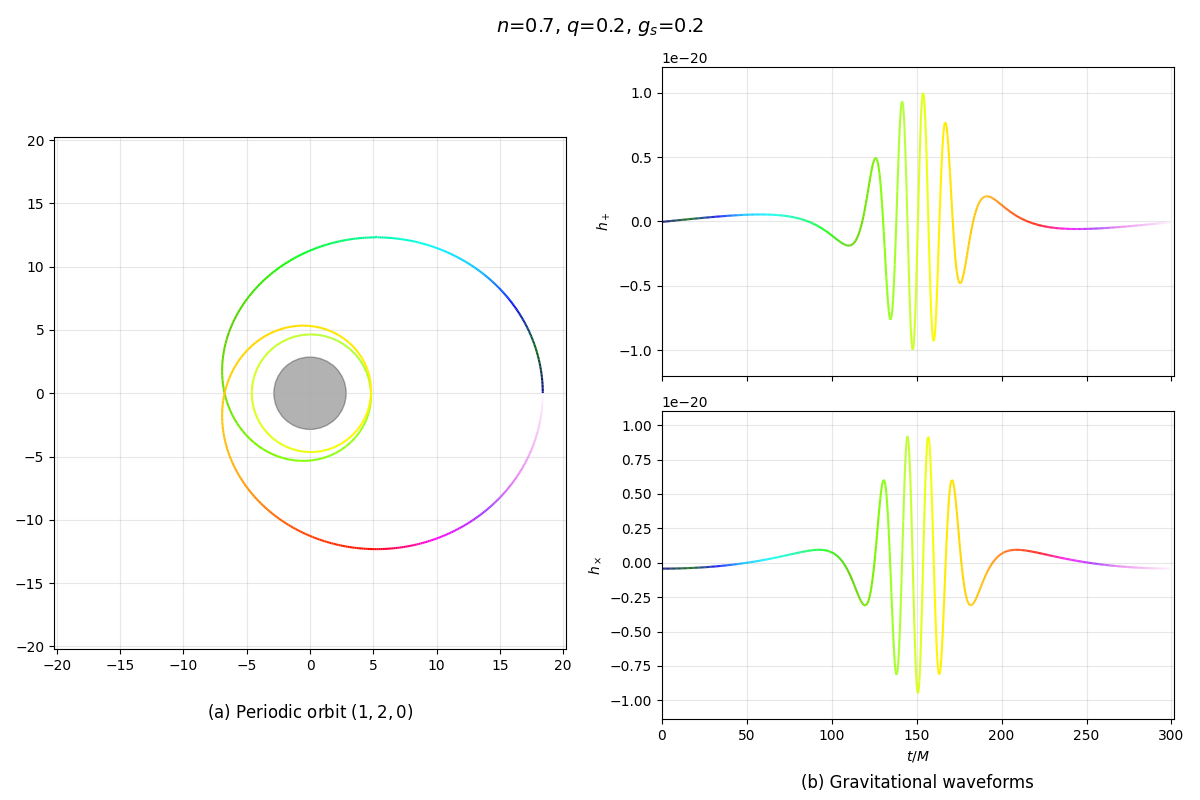}
\caption{Baseline gravitational wave strains ($h_+$ and $h_\times$) and corresponding spatial trajectory for a periodic orbit of class $(1, 2, 0)$ in the Freund-Nambu spacetime.}
\label{fig:waveforms_baseline}
\end{figure*}

\subsection{Physical Interpretation of Zoom-Whirl Kinematics}

The waveforms presented in Fig.~\ref{fig:waveforms_baseline} encapsulate the distinct dynamical fingerprints of relativistic zoom-whirl orbits. The periodic trajectories can be physically divided into two highly contrasting phases. During the \textit{zoom phase}, the particle moves through regions of relatively weak gravitational gradients, corresponding to extended intervals of low-frequency, low-amplitude gravitational wave emission. Conversely, during the \textit{whirl phase}, the particle plunges deep into the strong-field regime and grazes the unstable circular orbit. The extreme radial acceleration and relativistic velocities achieved in this deep potential well trigger intense, high-frequency bursts of gravitational radiation.

\subsection{Physical Interpretation of Scalar Parameter Dephasing}

The presence of the scalar field in the Freund-Nambu theory heavily modifies the near-horizon geometry and the particle's acceleration profile. The parameter $n$ controls the scalar modification to the JNW metric, $q$ governs the strength of the coupling between the scalar field and the geometry, and $g_s$ is the coupling constant between the scalar field and the massive particle.

To conduct a clean comparative analysis, we preserve the specific topological class of the orbit (specifically isolating the complex $(4,1,3)$ topology). To achieve this when tuning the scalar parameters, the orbital energy $E$ and angular momentum $L$ must be systematically adjusted. 

\begin{figure*}[htbp]
\centering
\includegraphics[width=0.9\textwidth]{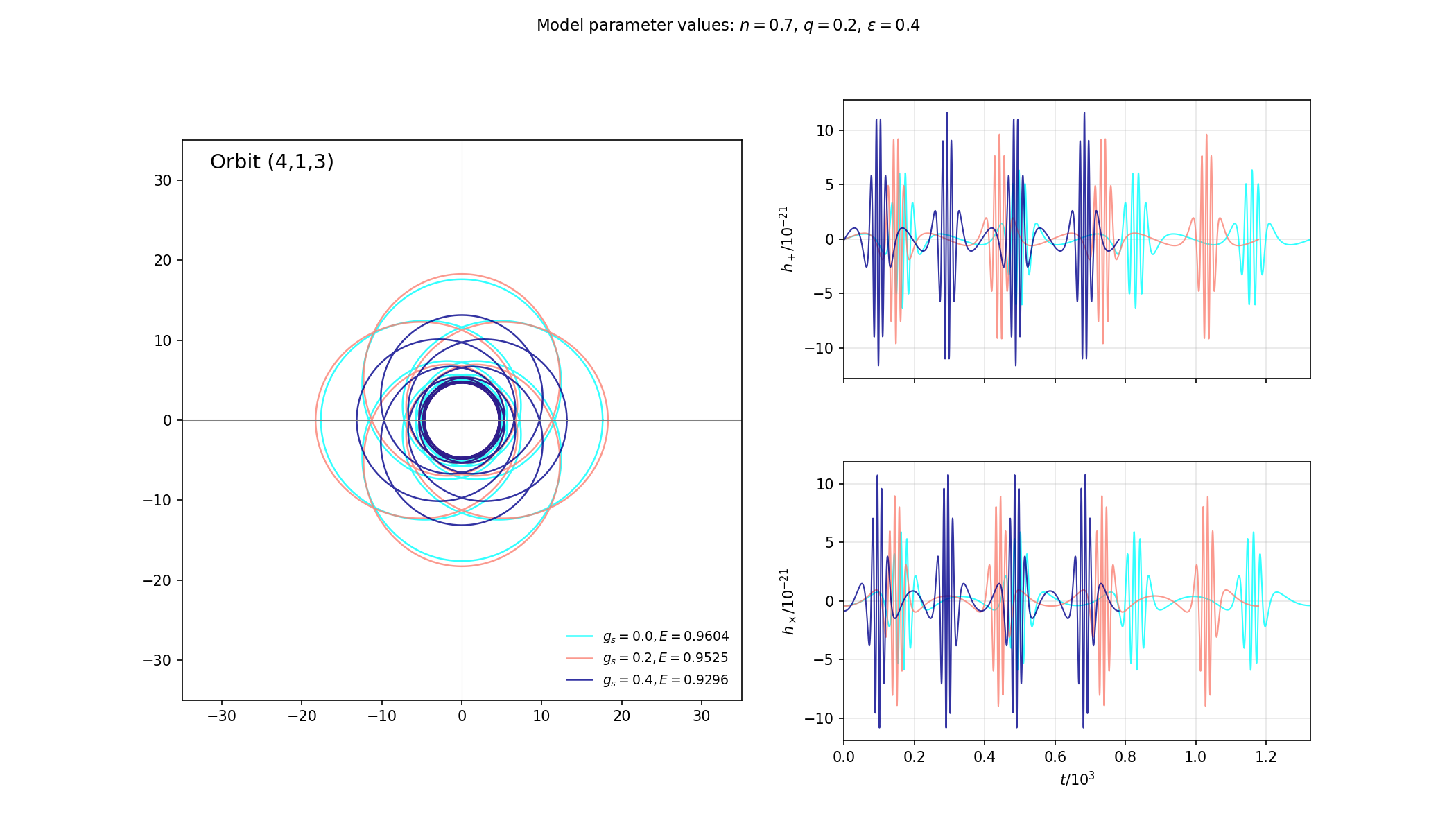}
\caption{The impact of the scalar-particle coupling parameter $g_s$ on EMRI waveforms for the $(4,1,3)$ topology. The variation induces severe temporal dephasing in the high-frequency whirl bursts.}
\label{fig:waveforms_gs}
\end{figure*}

\begin{figure*}[htbp]
\centering
\includegraphics[width=0.9\textwidth]{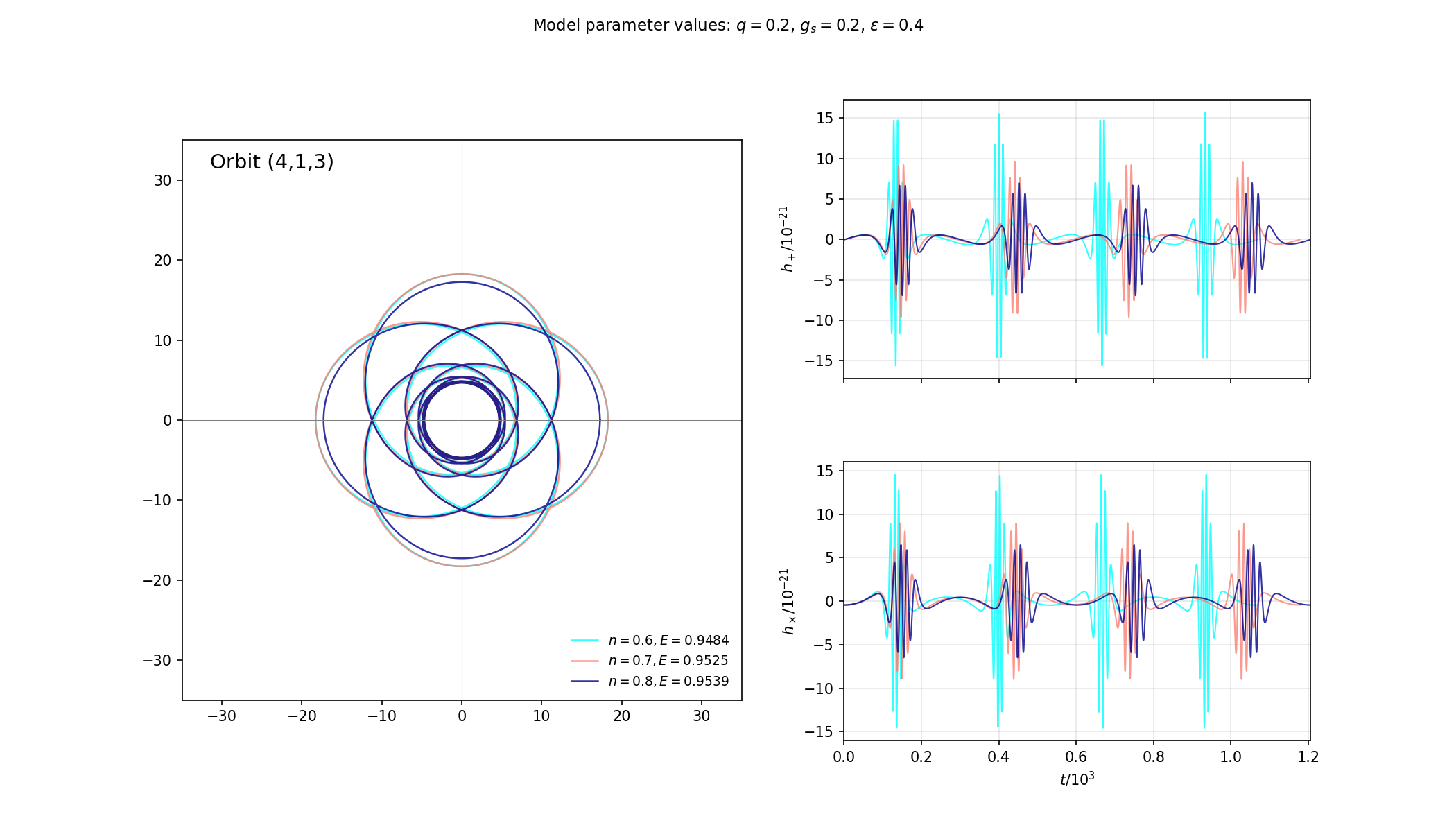}
\caption{The impact of the spacetime scalar parameter $n$ on EMRI waveforms for the $(4,1,3)$ topology.}
\label{fig:waveforms_n}
\end{figure*}

\begin{figure*}[htbp]
\centering
\includegraphics[width=0.9\textwidth]{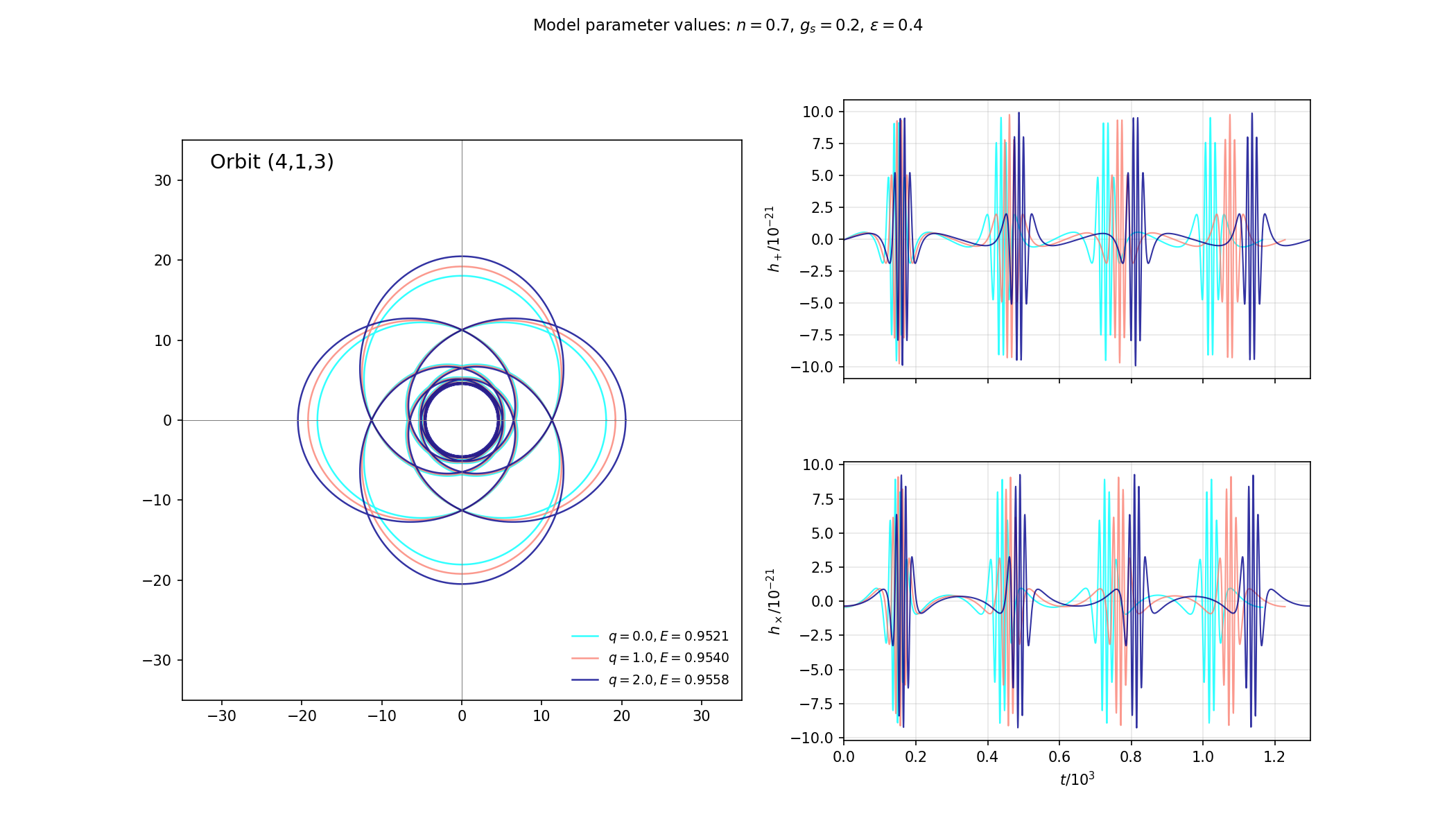}
\caption{The impact of the geometric scalar coupling $q$ on EMRI waveforms for the $(4,1,3)$ topology.}
\label{fig:waveforms_q}
\end{figure*}

Crucially, as shown in Figs.~\ref{fig:waveforms_gs}, \ref{fig:waveforms_n}, and \ref{fig:waveforms_q}, while the macroscopic spatial shapes of these tuned orbits look virtually identical, their time-domain evolution is fundamentally altered. Modifying $n$, $q$, or $g_s$ alters the steepness and shape of the effective radial potential $V(r)$. A steeper potential accelerates the particle differently as it falls into the whirl phase. Consequently, the proper time the particle spends traversing the strong-field region changes. 

This dynamic manifests in the waveforms as a severe and cumulative \textbf{macroscopic dephasing}. The arrival times of the intense high-frequency bursts shift systematically in the time domain as the scalar parameters are varied. This continuous phase accumulation proves that future space-based detectors can utilize these timing variations as precise theoretical templates to constrain the coupling parameters $q$ and $g_s$ alongside the scalar parameter $n$, effectively testing the Freund-Nambu gravity framework in the strong-field regime.

\section{Conclusions \label{Sec:Conclusions}}

In this paper, we have conducted a detailed and systematic investigation into the orbital dynamics and gravitational wave phenomenology of test particles within the framework of Freund-Nambu scalar-tensor gravity. Moving beyond the standard Schwarzschild paradigm, we focused on an exact, spherically symmetric vacuum solution that generalizes the well-known Janis-Newman-Winicour (JNW) spacetime. This geometry, which characterizes a naked singularity supported by a non-trivial scalar field profile, serves as a compelling theoretical laboratory for probing strong-field deviations from general relativity. By introducing a direct linear coupling parameter $g_s$ between the massive test particle and the background scalar field, alongside the geometric non-linear coupling $q$ and the spacetime parameter $n$, we established a robust analytical framework to evaluate how scalar hair dictates particle kinematics.

Utilizing the effective potential formalism, we rigorously mapped the parameter space governing stable and marginally bound equatorial orbits. Our analysis reveals that the complex interplay between $n$, $q$, and $g_s$ heavily distorts the effective potential barrier, systematically shifting the critical binding energies and the radial location of the Innermost Stable Circular Orbit (ISCO). Notably, we found that an attractive scalar-particle coupling ($g_s > 0$) pulls the ISCO deeper into the gravitational well, closer to the central naked singularity. This inward shift holds profound astrophysical implications, as it directly alters the inner edge of potential accretion disks and effectively enhances the radiative efficiency of accreting matter in these environments. Conversely, negative scalar couplings destabilize the orbits, pushing the ISCO outward and shrinking the permissible phase space for bounded oscillatory motion.

To probe the extreme strong-field regime, we cataloged families of bound periodic trajectories using the rational parameter $q_{\rm rat}$, establishing a precise $(z, w, v)$ taxonomy based on the number of zooms, whirls, and vertices. We demonstrated that as a particle's specific energy and angular momentum are meticulously tuned toward the separatrix limit, the scalar modifications induce intense periapsis precession, trapping the particle in pronounced zoom-whirl behavior. Because these complex, multi-petal orbits graze the unstable circular orbit, they are extraordinarily sensitive to the exact shape of the scalar-deformed potential barrier, serving as a powerful diagnostic tool for the near-singularity geometry.

Ultimately, the most striking observational consequences of these scalar-tensor modifications emerge in the gravitational wave sector. By applying the adiabatic approximation and the Numerical Kludge methodology with a linearized quadrupole emission model, we synthesized the gravitational wave strains emitted by Extreme Mass-Ratio Inspiral (EMRI) systems executing these periodic orbits. Our simulations clearly illustrate that the scalar-tensor corrections leave an indelible, macroscopic signature on the time-domain waveforms. Even when we rigorously preserve the spatial topology of an orbit by precisely tuning the orbital energy and angular momentum across different scalar parameters, the localized modifications to the gravitational field induce severe temporal dephasing during the high-frequency whirl bursts. 

This cumulative phase shift highlights the exquisite sensitivity of EMRI waveforms to the presence of scalar hair. The distinct temporal dephasing generated by $n$, $q$, and $g_s$ ensures that scalar-tensor modifications cannot simply mimic standard general relativistic effects without fundamentally altering the arrival times of the radiation bursts. In the era of precision gravitational wave astronomy, these unique temporal signatures will provide a robust theoretical template for future space-based interferometers, such as LISA. By capturing the rich phenomenology of zoom-whirl EMRIs, our results pave the way for utilizing future observational data to stringently constrain scalar-tensor extensions of Einstein's theory and to search for the definitive hallmarks of naked singularities in the cosmos.

\acknowledgments
Authors acknowledge useful discussions with Prof. Tao Zhu in preparing this manuscript.
The authors, DJG and JB, are grateful to the administration of Madhabdev University for their continued support and encouragement in carrying out this research.

\section*{Declaration of competing interest}
The authors declare that they have no known competing financial interests or personal relationships that could have appeared to influence the work reported in this manuscript.

\section*{Data Availability Statement}
There are no new data associated with this article.

\bibliographystyle{JHEP}
\bibliography{references}

\end{document}